\documentclass[aps,prb,twocolumn,showpacs,superscriptaddress,floatfix]{revtex4-1}
\usepackage{amssymb,amsmath}
\usepackage{graphicx}
\usepackage{dcolumn}
\usepackage{bm}
\usepackage{color}

\begin{document}
\title{Prediction of giant mechanocaloric effects in fluorite-structured superionic materials}
\author{Claudio Cazorla}
\thanks{Corresponding Author}
\affiliation{School of Materials Science and Engineering, UNSW Australia, Sydney NSW 2052, Australia \\
Integrated Materials Design Centre, UNSW Australia, Sydney NSW 2052, Australia}

\author{Daniel Errandonea}
\affiliation{Departamento de F\'isica Aplicada (ICMUV), Malta Consolider Team,
             Universitat de Valencia, 46100 Burjassot, Spain}

\begin{abstract}
Mechanocaloric materials experience a change in temperature when a mechanical stress is adiabatically
applied on them. Thus far, only ferroelectrics and superelastic metallic alloys have been considered 
as potential mechanocaloric compounds to be exploited in solid-state cooling applications. Here we show 
that giant mechanocaloric effects occur in hitherto overlooked fast ion conductors (FIC), a class of 
multicomponent materials in which above a critical temperature, $T_{s}$, a constituent ionic species 
undergoes a sudden increase in mobility. Using first-principles and molecular dynamics simulations, 
we found that the superionic transition in fluorite-structured FIC, which is characterised by a large 
entropy increase of the order of $10^{2}$~JK$^{-1}$Kg$^{-1}$, can be externally tuned with hydrostatic, 
biaxial or uniaxial stresses. In particular, $T_{s}$ can be reduced several hundreds of degrees through 
the application of moderate tensile stresses due to the concomitant drop in the formation energy of Frenkel 
pair defects. We predict that the adiabatic temperature change in CaF$_{2}$ and PbF$_{2}$, two archetypal 
fluorite-structured FIC, close to their critical points are of the order of $10^{2}$ and $10^{1}$~K, 
respectively. This work advocates that FIC constitute a new family of mechanocaloric materials  
showing great promise for prospective solid-state refrigeration applications.   
\end{abstract}
\maketitle

\begin{figure}[t]
\centerline
        {\includegraphics[width=1.0\linewidth]{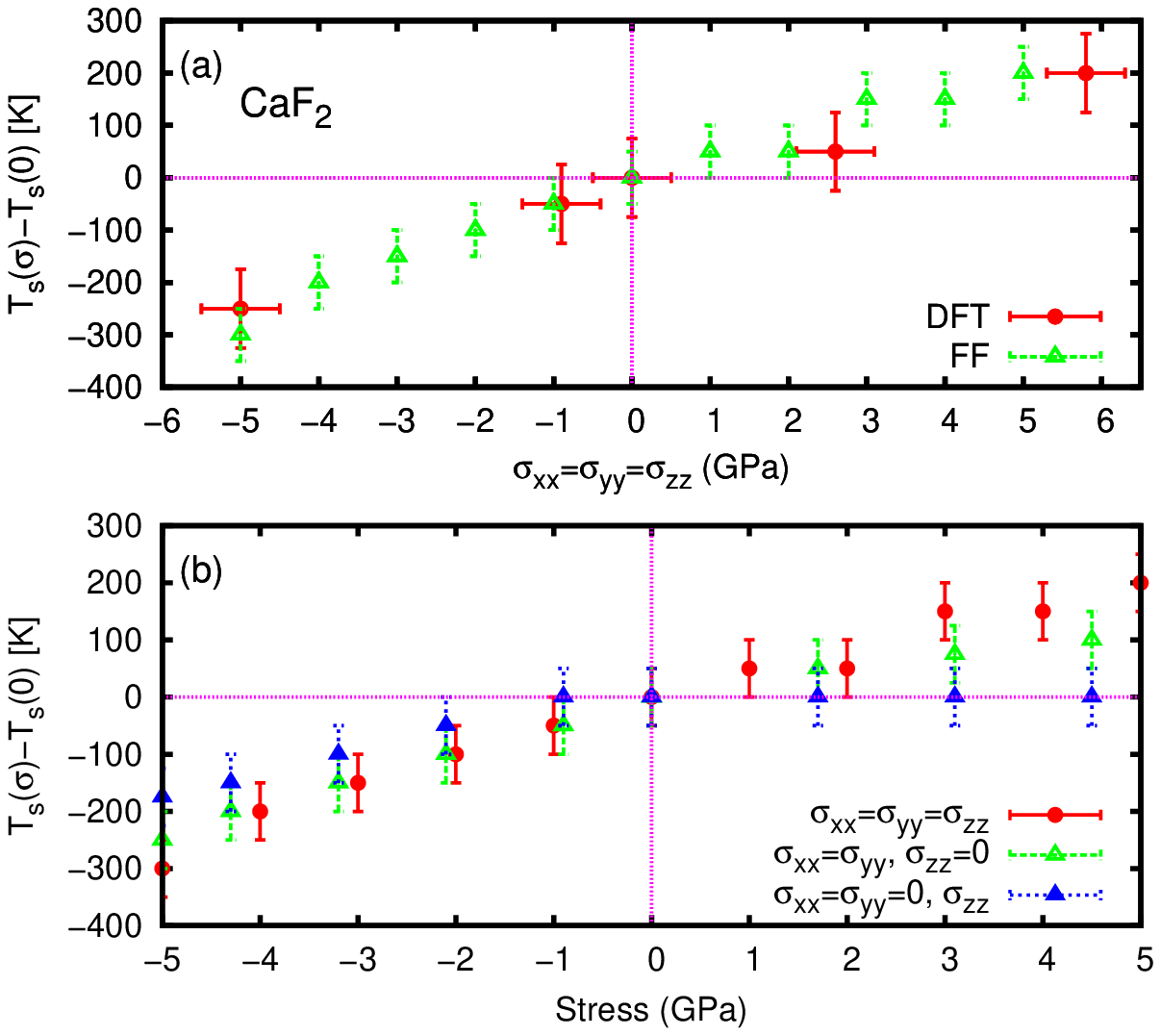}}
        {\includegraphics[width=0.9\linewidth]{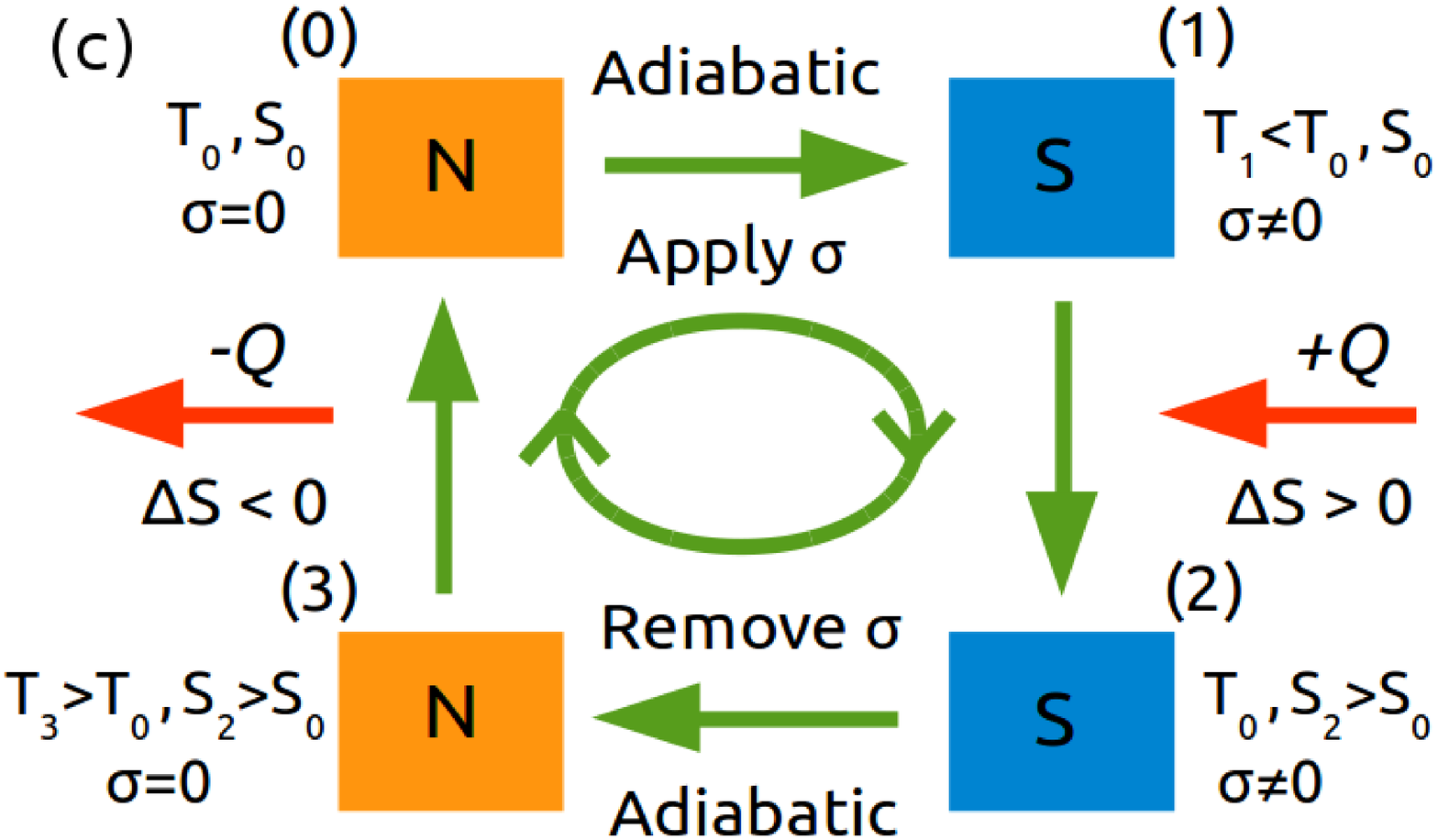}}
\caption{(a)~The superionic temperature in CaF$_{2}$ expressed as a function of hydrostatic
stress and calculated with first-principles (DFT) and molecular dynamics (FF) simulation methods.
(b)~Critical superionic temperature expressed as a function of hydrostatic, biaxial, and uniaxial 
stresses, calculated with molecular dynamics simulation methods.
(c)~Schematic representation of a mechanocaloric cooling cycle based on a fluorite-structured FIC at
$T_{0} < T_{s}(0)$. ``N'' and ``S'' represent the normal and superionic states. (0)~$\to$~(1)~A
tensile stress is adiabatically applied that triggers superionicity and thus the crystal gets cooler.
(1)~$\to$~(2)~The crystal receives heat from a system and thus $T$ and $S$ increase.
(2)~$\to$~(3)~The external stress is removed adiabatically and thus the crystal becomes normal
and $T$ increases. (3)~$\to$~(4)~Heat is ejected to the environment.}
\label{fig1}
\end{figure}

Fast-ion conductors (FIC) are solids in which ions are highly mobile. They are usually employed as electrolytes 
in solid-state batteries.~\cite{hayes85,hull04} 
Above a certain critical temperature, $T_{s}$, the anion or cation mobility in FIC becomes comparable to 
that of a molten salt, namely of the order of $1$~$\Omega^{-1}$cm$^{-1}$. This ``superionic'' transition 
can be thought of as a sublattice melting that, in analogy to homogeneous melting, has associated a large 
increase in entropy and lattice parameter.~\cite{andersen83,goff91} 
CaF$_{2}$ is an archetypal FIC that under ambient conditions crystallises in the cubic fluorite structure 
(space group $Fm\bar{3}m$). In this compound, the critical temperature for F$^{-}$ diffusivity is $1400(90)$~K 
and the accompanying raise in entropy $225.6$~JK$^{-1}$Kg$^{-1}$.~\cite{cazorla14,ubbelohde78} The accepted 
dominant effect behind the large ionic conductivity observed in CaF$_{2}$ and other analogous FIC is the 
formation of Frenkel pair defects (FPD), that is, the simultaneous creation of F$^{-}$ vacancies and 
interstitials.~\cite{gillan90,lindan93}

Recently, it has been demonstrated by state-of-the-art compression experiments and first-principles 
calculations that the superionic temperature in CaF$_{2}$ can be largely modified with the application of 
hydrostatic pressure. For instance, $T_{s}$ increases as much as $\sim 200$~K under a homogeneous load of 
$5$~GPa.~\cite{cazorla14} This fundamental finding suggests that external mechanical stress, $\sigma$, could 
be used to control the superionic transition in FIC, a possibility that, due to the huge entropy change 
associated to the transformation and structural simplicity and abundance of the involved materials, could be 
highly exploitable in energy conversion applications. However, a thorough understanding of the atomic mechanisms 
mediating the observed stress-induced $T_{s}$ variation is still lacking, and thus possible scientific 
and technological developments are being hampered. Here, we apply a fully atomistic simulation approach to 
fill this critical knowledge gap. In particular, we rationalise how the critical temperature in fluorite-structured 
FIC is affected by compressive ($\sigma > 0$) and tensile ($\sigma < 0$) hydrostatic, biaxial and uniaxial 
stresses, and evaluate the potential of this effect for solid-state cooling operation.         

Three types of mechanical stress were considered in our study: hydrostatic ($\sigma_{xx} = 
\sigma_{yy} = \sigma_{zz}$), biaxial ($\sigma_{xx} = \sigma_{yy}, \sigma_{zz} = 0$), and uniaxial
($\sigma_{xx} = \sigma_{yy} = 0, \sigma_{zz}$). We adopted a rigid-ion Born-Mayer-Huggins (BMH) 
interatomic potential to describe the interactions between atoms in CaF$_{2}$.~\cite{cazorla13,cazorla15} 
This interaction potential renders a satisfactory description of $T_{s}$ under varying hydrostatic 
stress, as it is demonstrated in Fig.~\ref{fig1}a through the comparison to first-principles results 
obtained with density functional theory (DFT) [for technical details, see Supplementary Information].
At equilibrium ($\sigma = 0$), the adopted BMH potential provides a superionic critical temperature 
of $1350(50)$~K, which is in good accordance with reported experimental data and DFT calculations.~\cite{cazorla14} 
In view of such an agreement, we assume that the classical BMH and first-principles DFT results obtained 
in the rest of cases are also consistent.     

Figure~\ref{fig1}b shows the stress dependence of $T_{s}$ calculated under broad $\sigma > 0$ (compressive) 
and $\sigma < 0$ (tensile) conditions. At compressive stress, the superionic features in CaF$_{2}$ rely  
markedly on the type of $\sigma$ that is applied. For instance, in the hydrostatic case $T_{s}$ increases 
as much as $\sim 200$~K under a maximum load of $5$~GPa whereas the same critical temperature remains 
practically insensitive to uniaxial compressive stresses of the same magnitude. The results obtained for 
biaxial compressive stresses shows a tendency that is kind of an average between the hydrostatic and uniaxial 
cases. Under $\sigma < 0$ conditions, however, all three types of stresses produce similar effects on CaF$_{2}$ 
and superionicity emerges at temperatures significantly lower than at equilibrium. For instance, at the 
maximum tensile load considered here $T_{s}$ is reduced as much as $200-300$~K. 
             
In view of the $T_{s}(\sigma)$ results presented in Fig.~\ref{fig1}b, one can think of original mechanocaloric 
cooling cycles involving FIC. Among all the possibilities, we sketch in Fig.~\ref{fig1}c one consisting of two 
adiabatic and two constant uniaxial $\sigma < 0$ steps. The change in temperature that occurs during the two 
adiabatic processes is 
\begin{equation}
\Delta T = - \int_{0}^{\sigma} \frac{T}{C_{\sigma}} \cdot dS = - \int_{0}^{\sigma} \frac{T}{\rho C_{\sigma}} \cdot \left( \frac{\partial \epsilon}{\partial T} \right)_{\sigma} d\sigma~, 
\label{eq:adiaT}
\end{equation}
where $\rho$, $C_{\sigma}$, and $\epsilon$ represent the density, heat capacity and mechanical strain in 
the FIC. The operation temperature for such an hypothetical refrigeration cycle is $T < T_{s}(0)$, that 
is, the FIC is initialised from the nonsuperionic (or normal) state. Upon application~(removal) of tensile 
stress the entropy of the crystal increases~(decreases) due to triggering~(prevention) of the superionic 
state, thereby its temperature decreases~(increases) [that is, $\Delta S > 0$ implies $\Delta T < 0$ and 
viceversa, see Eq.~\ref{eq:adiaT}]. We note that the represented cooling sequence works in ``inverse'' 
order to usual refrigeration cycles based on ferroelectrics and shape-memory alloys.~\cite{scott11,liu14,tusek15,tusek15b} 
The reason for this is that the state of maximum entropy is accessed through the switching-on of the external 
field. Nonetheless, the normal order can be recovered by setting $T > T_{s}(0)$ operation conditions and applying 
compressive stresses.   

\begin{figure}[t]
\centerline
        {\includegraphics[width=1.0\linewidth]{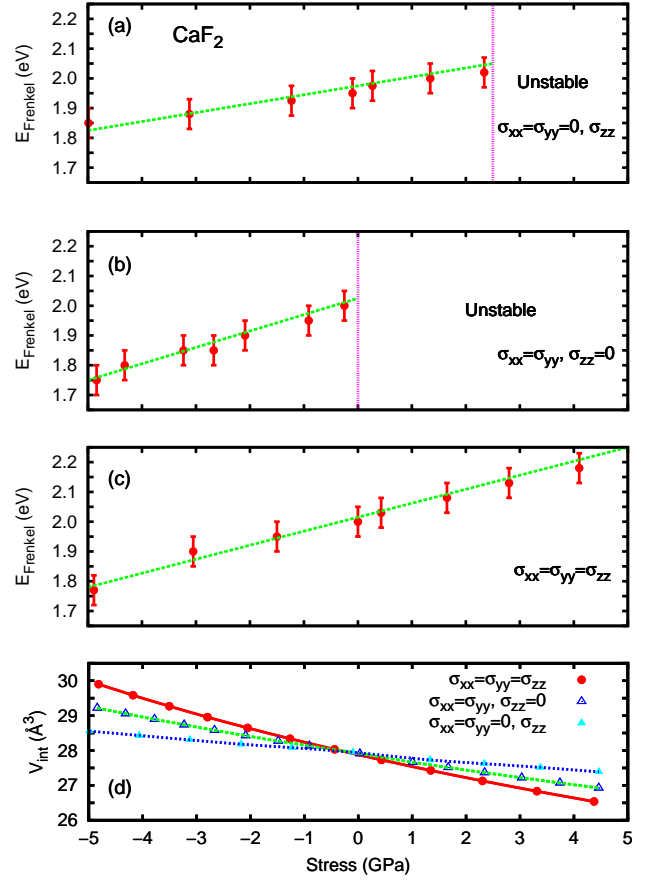}}
\caption{Formation energy of Frenkel pair defects in CaF$_{2}$ calculated with first-principles 
DFT methods and expressed as a function of uniaxial~(a), biaxial~(b), and hydrostatic~(c) stress. 
``Unstable'' indicates that during relaxation of the crystal the F$^{-}$ interstial  
returned to its equilibrium lattice position. (d)~Available volume to the interstitial  
expressed as a function of stress.}                               
\label{fig2}
\end{figure}

In order to rationalise the physical origins of the $T_{s}(\sigma)$ results shown in Fig.~\ref{fig1},
we computed the formation energy of Frenkel pair defects (FPD) for all considered $\sigma$ with 
first-principles DFT methods (see Fig.~\ref{fig2}). For this, we constructed a cubic supercell containing 
$144$ atoms that subsequently was relaxed according to the imposed stress conditions. (Structural phase 
transitions were absent in all the performed geometry optimizations [see Supplementary Information].) 
An arbitrary F$^{-}$ ion then was moved away from its equilibrium position according to the Cartesian 
displacement $(u, v, w)$, where $0 \le u, v, w \le \frac{a_{i}}{2}$ and $\lbrace a_{i} \rbrace$ are the 
lattice parameters of the corresponding unit cell. The three Cartesian direction were sampled with $5$ 
equidistant starting points (that is, a total of $125$ relaxations were performed in each case). We 
systematically found that the only metastable interstitial configuration associated to the fluorine 
position was $(\frac{a_{x}}{2}, \frac{a_{y}}{2} , \frac{a_{z}}{2})$, which ordinarily is known as the 
octahedral site. As it is shown in Figs.~\ref{fig2}a and b, this configuration became also unstable 
(that is, the F$^{-}$ ion returned to its equilibrium lattice site during the relaxation) under moderate 
uniaxial and biaxial stresses. We note that even though other interstitial positions are likely to be 
stabilised by effect of temperature,~\cite{hull04,castiglione01} we disregarded thermal excitations in 
this part of our study.        

Our results in Fig.~\ref{fig2} show that the overall effect of applying compressive stress in CaF$_{2}$ is to
increase the formation energy of FPD, thereby hindering superionicity. On the contrary, tensile stress clearly
enhances F$^{-}$ mobility by depleting the corresponding migration energy barrier. The variation of the FPD formation 
energy behaves linearly with respect to the external stress and is not bounded from below. These trends are correlated 
with the variation of the volume that is available to the fluorine interstitial, $V_{\rm int}$, which is defined 
as the empty octahedron space in the perfect fluorite structure (see Fig.~\ref{fig2}d).~\cite{hull04,castiglione01} 
This volume expands or reduces roughly in proportion to $\sigma$, depending on whether the applied stress 
is tensile or compressive. In analogy to $T_{s}$, hydrostatic stress induces the largest $V_{\rm int}$ variation 
whereas uniaxial stress the smallest. Also, the larger $V_{\rm int}$ is the smaller the FPD formation energy results. 
These structural and migration energy barrier outcomes clarify the causes behind the $T_{s}(\sigma)$ 
trends shown in Fig.~\ref{fig1}, confirming hydrostatic and biaxial stresses as most effective for 
tuning of the transport properties in FIC. (These findings may have an immediate application in the design of 
improved electrochemical devices however, for the sake of focus, we do not elaborate on this aspect here.) 

We note that in practice tensile stress can be achieved both in the uniaxial and biaxial cases. Actually, CaF$_{2}$ 
can be deposited as a thin film on different substrates.~\cite{maki02,pilvi07,pandey14} Tensile stresses are induced 
by the in-plane lattice mismatch between CaF$_{2}$ and the substrate, when the latter has a larger lattice parameter.
According to our first-principles DFT calculations, $\sigma_{xx} = \sigma_{yy} = -5$~GPa conditions, for instance, 
correspond to an epitaxial strain of $\eta = +2.8$~\% (where $\eta \equiv a - a_{0}/a_{0}$ and $a_{0} = 5.52$~\AA).  
In this context, Germanium appears to be a good candidate substrate since it is structurally compatible with CaF$_{2}$ 
and has a lattice parameter of $\sim 5.7$~\AA.  

\begin{figure*}[t]
{\includegraphics[width=1.0\linewidth]{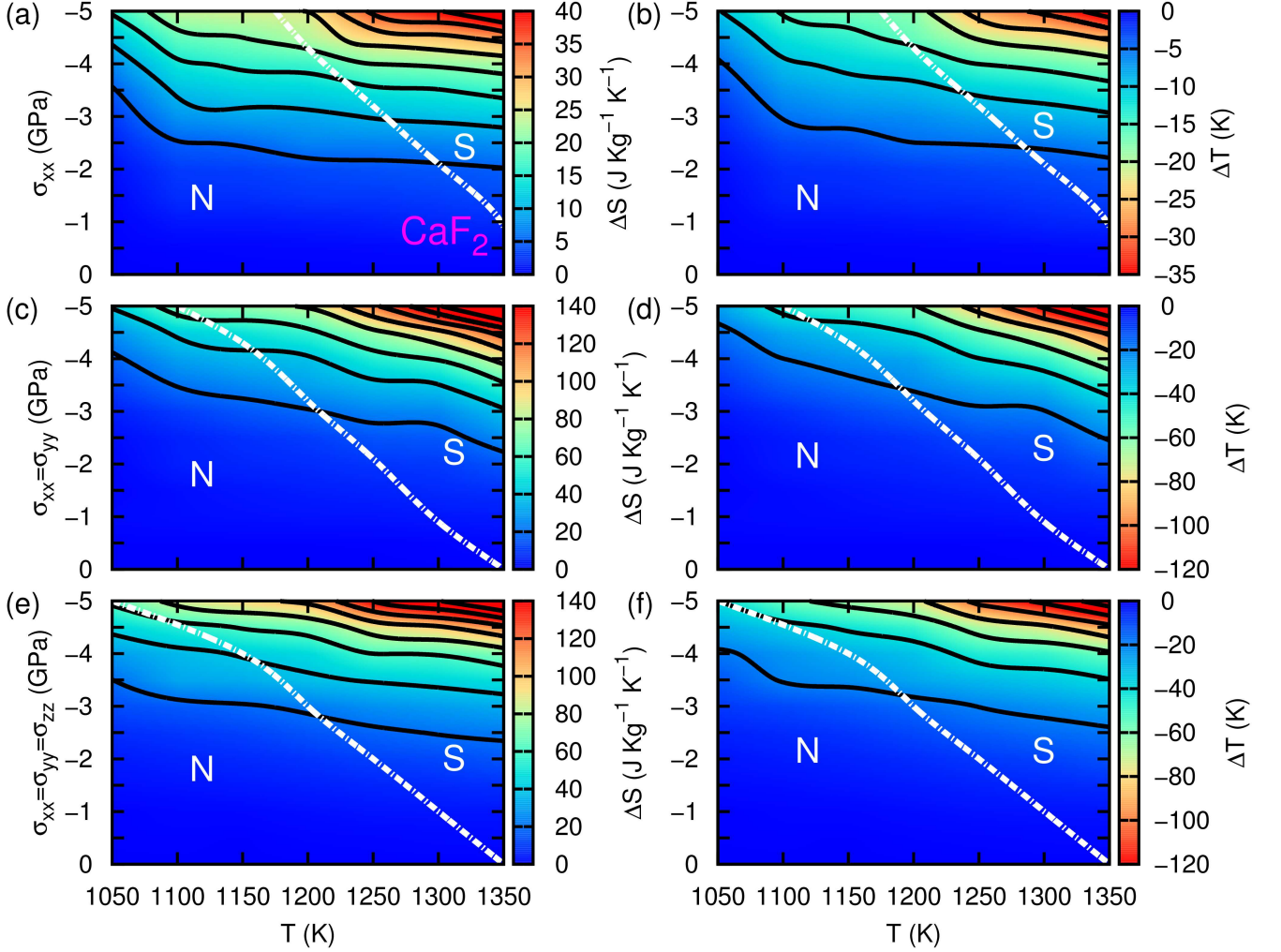}}
\caption{Isothermal entropy, $\Delta S$, and adiabatic temperature, $\Delta T$, changes in CaF$_{2}$ 
calculated with molecular dynamics simulation techniques and expressed a function of stress
and temperature. Results enclosed in (a)-(b) correspond to uniaxial stress, (c)-(d) to biaxial, 
and (e)-(f) to hydrostatic. ``N'' and ``S'' represent the normal and superionic states and the 
thick dashed lines mark the corresponding phase boundaries.}
\label{fig3}
\end{figure*}

In order to assess the potential of fluorite-structured FIC for prospective solid-state cooling applications, 
we calculated the adiabatic temperature change induced on CaF$_{2}$ by the application of tensile stress, 
$\Delta T$, with molecular dynamics simulations (i. e., using the BMH interaction model). To this end, we computed 
the isothermal entropy change associated to each type of stress, $\Delta S$, and heat capacity of the crystal 
as a function of $T$ and $\sigma$, and subsequently integrated them according to Eq.~(\ref{eq:adiaT})
or a similar expression [for technical details, see Supplementary Information].~\cite{moya14} It is worth 
stressing that the only approximations affecting to our results are referred to the BMH potential, which 
otherwise it has been demonstrated to be accurate enough for present purposes. Neither phenomenological 
models nor experimental data were assumed in our calculations. 

The computed $\Delta S$ and $\Delta T$ considering a maximum tensile stress of $5$~GPa and 
temperatures $T_{s}(0)-300~{\rm K} \le T \le T_{s}(0)$, are shown in Fig.~\ref{fig3}. The results obtained 
in the three $\sigma$ cases are qualitatively very similar. At $\sigma = 0$ conditions, for example, both 
the isothermal entropy and adiabatic temperature changes are practically null. This applies even to the 
highest analysed temperature because we identified $T_{s}$ with the onset of F$^{-}$ diffusivity, instead 
of the sudden increase in heat capacity that occurs at higher $T$ when superionicity is fully 
developed [see Supplementary Information].~\cite{andersen83,goff91} As tensile stress is raised, both $\Delta S$ 
and $\Delta T$ steadily increase in absolute value and their variation becomes larger at higher temperatures. 
As a consequence, no stationary points were found in our adiabatic or isothermal calculations in consistency  
with the findings shown in Fig.~\ref{fig2}. From a quantitative point of view, the results obtained in the 
hydrostatic and biaxial cases are both comparable and superior in terms of mechanocaloric potential to those 
found for uniaxial stresses. For instance, at $T = 1350$~K and considering $\sigma = -5$~GPa the adiabatic 
temperature change~(istohermal entropy change) calculated in the hydrostatic, biaxial and uniaxial 
cases are $\sim -152$~($186$), $-163$~($200$), and $-38$~K~($46$~JK$^{-1}$Kg$^{-1}$), respectively.   

The mechanocaloric results just presented reveal that fluorite-structured FIC are auspicious materials for 
solid-state refrigeration. In spite of the fact that the involved temperatures are well above ambient conditions 
and that the considered tensile stresses are moderately large, considering FIC in cooling applications may result 
in several advantages with respect to usual ferroelastic and ferroic materials. First, the predicted $\Delta T$ 
and $\Delta S$ are about one order of magnitude larger than (hydrostatic case) or comparable to (uniaxial case) 
the benchmark results reported thus far.~\cite{moya14,manosa13} Second, the analysed normal~$\to$~superionic 
transition is of second-order type and, in contrast to ferroic materials for example, FIC do not present 
order-parameter domains. These features are highly desirable for improved cyclability and rate capability of 
likely cooling devices, since thermal and mechanical hysteresis effects deriving from irreversible processes 
then would be littlest.~\cite{tusek15b,santana11} Actually, recent ultrafast X-ray spectroscopy experiments have 
demonstrated that the characteristic timescale of superionic switching is of the order of few picoseconds.~\cite{miller13} 
And third, the predicted $\Delta T$ and $\Delta S$ exhibit a uninterrupted escalation with respect to tensile stress 
(see Fig.~\ref{fig2}). This means that one could virtually go down to the ideal tensile strength of the crystal, 
which corresponds to its mechanical instability limit, in order to maximally lower $T_{s}$ and augment $|\Delta T|$. 
We note that in ferroelectric-paraelectric or austenite-martensite transformations, the resulting adiabatic 
temperature changes inevitably start decreasing beyond a certain threshold value of the external field due to 
saturation of the involved degrees of freedom.~\cite{liu14,tusek15}

\begin{figure}[t]
\centerline
        {\includegraphics[width=1.0\linewidth]{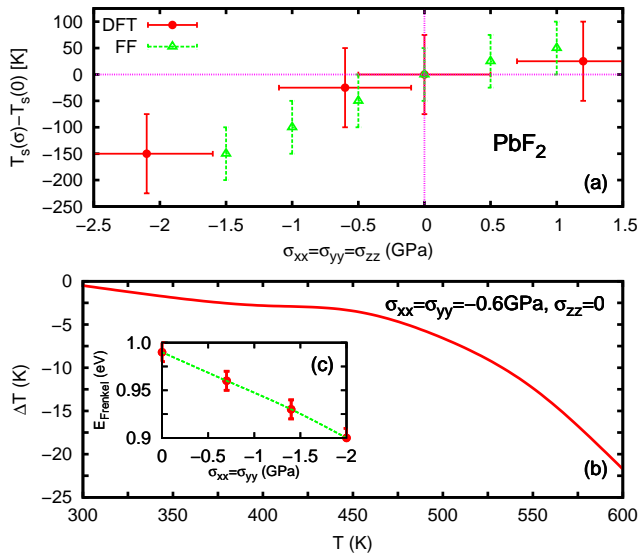}}
\caption{(a)~The superionic temperature in PbF$_{2}$ expressed as a function of hydrostatic
stress and calculated with first-principles (DFT) and molecular dynamics (FF) simulation methods.
(b)~Adiabatic temperature change, $\Delta T$, calculated in PbF$_{2}$ with molecular 
dynamics simulation methods and expressed as a function of temperature. The applied biaxial tensile
stress is $0.6$~GPa. (c)~Formation energy of Frenkel pair defects in PbF$_{2}$ calculated with 
first-principles DFT methods and expressed as a function of biaxial stress.}
\label{fig4}
\end{figure}

In fact, the critical temperature in superionic materials can be reduced significantly by means of 
nanopatterning and chemical substitution strategies.~\cite{hull04,makiura09} Aimed at alleviating 
the technical shortcomings found in CaF$_{2}$, we investigated the same class of superionic and mechanocaloric 
phenomena in PbF$_{2}$, a related fluorite-structured FIC with a much lower transition temperature of 
$T_{s}(0) \sim 700$~K.~\cite{schroter80} (We note that by minimally doping PbF$_{2}$ with potassium ions
it is possible to reduce the corresponding critical temperature practically down to ambient.~\cite{hull98})  
We adopted a rigid-ion BMH interatomic potential to describe the interactions between atoms in PbF$_{2}$.~\cite{walker82} 
This interaction potential also renders a satisfactory description of $T_{s}$ under varying hydrostatic stress, 
as it is demonstrated in Fig.~\ref{fig4}a through the comparison to first-principles DFT results. At equilibrium 
the adopted BMH potential provides a superionic critical temperature of $650(50)$~K, which is in good agreement 
with experiments and DFT calculations.

The $T_{s}(\sigma)$ trends in PbF$_{2}$ under hydrostatic and biaxial stresses are qualitatively 
analogous to those found in CaF$_{2}$ (see Figs.~\ref{fig4}a,b). Namely, tensile stress effectively
depletes the corresponding critical temperature whereas compressive stress increases it. Also, the energy 
barrier for F$^{-}$ migration decreases roughly in proportion to the tensile stress and is not disrupted 
from below (see Fig.~\ref{fig4}c). We note that in our classical BMH and first-principles DFT simulations 
it was not possible to reproduce hydrostatic stress conditions below $\sim -3$~GPa due to mechanical 
instabilities that started to develop in the system. We tentatively identify such a stress threshold with the 
ideal tensile strength in PbF$_{2}$ (a similar regime was accessed also in CaF$_{2}$ although at $\sigma < -6$~GPa). 
At the quantitave level, we found two principal differences between the two investigated FIC. 
First, smaller stresses are needed in PbF$_{2}$ to achieve a same critical temperature reduction. For example, 
a $T_{s}(\sigma) - T_{s}(0)$ difference of $-150$~K is produced by a hydrostatic stress of $\sigma \sim -1.5 $~GPa in 
PbF$_{2}$ and of $\sim -3.0$~GPa in CaF$_{2}$. And second, when considering a same tensile stress the adiabatic 
temperature change calculated close to the critical point is larger (in absolute value) in PbF$_{2}$. For instance, 
at $\sigma_{xx} = \sigma_{yy} = -0.6$~GPa the estimated $\Delta T$ in PbF$_{2}$ is $\sim -20$~K (see Fig.~\ref{fig4}b) 
while in CaF$_{2}$ $\sim -2$~K (see Fig.~\ref{fig3}d). This finding can be rationalised in terms of the accompanying 
isothermal entropy changes, which for small $\sigma$ are much larger in PbF$_{2}$. Finally, we note that according 
to our first-principles DFT calculations biaxial tensile stress of $\sim -0.5$~GPa are realisable in PbF$_{2}$ thin 
films with a small epitaxial strain of $\eta = +0.5$~\% (where $a_{0} = 6.00$~\AA). The results just explained 
indicate that PbF$_{2}$ is a promising material for near-room-temperature cooling applications. 

In summary, we have employed classical molecular dynamics and first-principles DFT simulation techniques to 
discern the relations between external mechanical stress and ionic transport in fluorite-structured FIC. 
Our computational study shows that hydrostatic, biaxial, and uniaxial stresses can be used as effective means 
for tuning the critical temperature in superionic compounds. This finding may have important implications
for the design of solid-state batteries with improved ion diffusion kinetics. We have predicted that 
the adiabatic temperature change occurring in fluorite-structured FIC under external tensile stress, is comparable 
in magnitude to current benchmark results reported for ferroelastic and ferroic materials. Our conclusions for 
CaF$_{2}$ and PbF$_{2}$ could be generalised to other FIC, like for instance Ag$^{+}$ chalcogenides and halides 
and Li-based complex hydrides, with critical temperatures closer to room temperature. The present work, therefore, 
opens a new and promising avenue for rational design of original refrigeration materials.

\section*{Methods}
${\bf Classical~and~DFT~computer~simulations.}$~Molecular dynamics $( N, P, T )$ simulations 
were performed with the LAMMPS code.~\cite{lammps} The pressure and temperature in the system were kept 
fluctuating around a set-point value by using thermostatting and barostatting techniques in which some 
dynamic variables are coupled to the particle velocities and simulation box dimensions. Large simulation 
boxes containing $6,144$ atoms were used and periodic boundary conditions were applied along the three 
Cartesian directions. Newton's equations of motion were integrated using the customary Verlet's algorithm 
with a time-step length of $10^{-3}$~ps. A particle-particle particle-mesh $k$-space solver was used to 
compute long-range van der Waals and Coulomb interactions and forces beyond a cut-off distance of $12$~\AA~ 
at each time step.

First-principles DFT calculations were performed with the VASP code,~\cite{vasp} following the generalized 
gradient approximation to the exchange-correlation energy due to Perdew~\cite{pbe96}. The ``projector 
augmented wave'' method was used to represent the ionic cores~\cite{bloch94}, and Ca's $2s$-$3s$-$3p$-$4s$, 
Pb's $5d$-$6s$-$6p$ and F's $2s$-$2p$ electronic states were considered as valence. Wave functions were 
represented in a plane-wave basis truncated at $500$~eV. By using these parameters and dense ${\bf k}$-point 
grids for Brillouin zone integration, the resulting enthalpies were converged to within $1$~meV per formula 
unit. In the geometry relaxations, a tolerance of $0.01$~eV$\cdot$\AA$^{-1}$ was imposed in the atomic forces. 
Further details of our classical and \emph{ab initio} molecular dynamics simulations can be found in the 
Supplementary Information.

\bigskip

\section*{Acknowledgments}
This research was supported under the Australian Research Council's Future Fellowship 
funding scheme (project number FT140100135). Computational resources and technical assistance 
were provided by the Australian Government through Magnus under the National Computational Merit 
Allocation Scheme. DE acknowledges financial support from Spanish MINECO under Grants 
MAT2013-46649-C04-01/02/03 and MAT2015-71070-REDC (MALTA Consolider).

\section*{Author contributions}
All authors contributed equally to the present work.

\section*{Additional information}
Supplementary information is available in the online version of the paper.

\section*{Competing financial interests}
The authors declare no competing financial interests.


\begin{thebibliography}{30}
\bibitem{hayes85} W. Hayes and A. M. Stoneham in \textit{Defects and Defect Processes in Non-metallic Solids},
                  (Wiley, New York 1985).
\bibitem{hull04} S. Hull, Rep. Prog. Phys. \textbf{67}, 1233 (2004).  
\bibitem{andersen83} N. H. Andersen, K. Clausen, and J. K. Kjems, Sol. Stat. Ion. \textbf{9}, 543 (1983).
\bibitem{goff91} J. P. Goff, W. Hayes, S. Hull, and M. T. Hutchings, J. Phys.: Condens. Matt. \textbf{3}, 3677 (1991).
\bibitem{cazorla14} C. Cazorla and D. Errandonea, Phys. Rev. Lett. \textbf{113}, 235902 (2014).
\bibitem{ubbelohde78} A. R. Ubbelohde in \textit{The Molten State of Matter}, (Wiley, New York 1978).
\bibitem{gillan90} M. J. Gillan, J. Phys. C \textbf{19}, 3391 (1986);
                   J. Chem. Soc. Faraday Trans. \textbf{86}, 1177 (1990).
\bibitem{lindan93} P. J. D. Lindan and M. J. Gillan, J. Phys.: Condens. Matt. \textbf{5}, 1019 (1993).
\bibitem{cazorla13} C. Cazorla and D. Errandonea, J. Phys. Chem. C \textbf{117}, 11292 (2013).
\bibitem{cazorla15} C. Cazorla, Res. Phys. \textbf{5}, 262 (2015).
\bibitem{scott11} J. F. Scott, Annu. Rev. Mater. Res. \textbf{41}, 229 (2011).
\bibitem{liu14} Y. Liu, J. Wei, P.-E. Janolin, I. C. Infante, J. Kreisel, X. Lou, and B. Dkhil, 
                Phys. Rev. B \textbf{90}, 104107 (2014). 
\bibitem{tusek15} J. Tu${\rm \check{s}}$ek, K. Engelbrecht, R. Mill\'{a}n-Solsona, L. Ma${\rm \tilde{n}}$osa, E. Vives, L. P. Mikkelsen, and N. Pryds, 
                  Adv. Energy Mater. \textbf{5}, 1500361 (2015).
\bibitem{tusek15b} J. Tu${\rm \check{s}}$ek, K. Engelbrecht, L. P. Mikkelsen, and N. Pryds, J. Appl. Phys. \textbf{117}, 124901 (2015).
\bibitem{castiglione01} M. J. Castiglione and P. A. Madden, J. Phys.: Condens. Matt. \textbf{13}, 9963 (2001).
\bibitem{maki02} T. Maki, K. Okamoto, M. Sugiura, T. Hosomi, and T. Kobyashi, Appl. Surf. Sci. \textbf{197}, 448 (2002).
\bibitem{pilvi07} T. Pilvi, K. Arstila, M. Leskel\"{a}, and M. Ritala, Chem. Mater. \textbf{19}, 3387 (2007). 
\bibitem{pandey14} R. K. Pandey, M. Kumar, S. A. Khan, T. Kumar, A. Tripathi, D. K. Avasthi, and A. C. Pandeya, 
                   Appl. Surf. Sci. \textbf{289}, 77 (2014). 
\bibitem{moya14} X. Moya, S. Kar-Narayan, and N. D. Mathur, Nature Mat. \textbf{13}, 439 (2014).
\bibitem{manosa13} L. Ma${\rm \tilde{n}}$osa, A. Planes, and M. Acet, J. Mater. Chem. A \textbf{1}, 4925 (2013).
\bibitem{santana11} R. P. Santana, N. A. de Oliveira, and P. J. von Ranke, J. All. Comp. \textbf{509}, 6346 (2011).
\bibitem{miller13} T. A. Miller, J. S. Wittenberg, H. Wen, S. Connor, Y. Cui, and A. M. Lindenberg, Nat. Comm. \textbf{4}, 1369 (2013).
\bibitem{makiura09} R. Makiura, T. Yonemura, T. Yamada, M. Yamauchi, R. Ikeda, H. Kitagawa, K. Kato, and M. Takata, 
                    Nat. Mater. \textbf{8}, 476 (2009).
\bibitem{schroter80} W. Schr\"{o}ter and J. Nolting, J. Phys. Coll. \textbf{41}, 20 (1980).
\bibitem{hull98} S. Hull, P. Berastegui, S. G. Eriksson, and N. J. G. Gardner, J. Phys.: Condens. Matt. \textbf{10}, 8429 (1998).
\bibitem{walker82} A. B. Walker, M. Dixon, and M. J. Gillan, J. Phys. C: Sol. Stat. Phys., \textbf{15}, 4061 (1982). 
\bibitem{lammps} S. J. Plimpton, J. Comp. Phys. \textbf{117}, 1 (1995); \emph{http://lammps.sandia.gov}.
\bibitem{vasp} G. Kresse and J. Furthm\"{u}ller, Phys. Rev. B \textbf{54}, 11169 (1996). 
\bibitem{pbe96}	 J. P. Perdew, K. Burke, and M. Ernzerhof, Phys. Rev. Lett. \textbf{77}, 3865 (1996).
\bibitem{bloch94} P. E. Bl\"ochl, Phys. Rev. B \textbf{50}, 17953 (1994).
\end{thebibliography}
\end{document}